# Spectral Decomposition of Thermal Conductivity: Comparing Velocity Decomposition Methods in Homogeneous Molecular Dynamics Simulations

Alexander J. Gabourie[1], Zheyong Fan[2,3], Tapio Ala-Nissilä[2,4], and Eric Pop[1,5,6,*]

[1]Department of Electrical Engineering, Stanford University, Stanford, CA 94305, USA

[2]MSP group, QTF Centre of Excellence, Department of Applied Physics, Aalto University, FI-00076 Aalto, Finland

[3]College of Physical Science and Technology, Bohai University, Jinzhou, 121013, China

[4]Interdisciplinary Centre for Mathematical Modelling, Department of Mathematical Sciences, Loughborough University, Loughborough, Leicestershire LE11 3TU, UK

[5]Department of Materials Science & Engineering, Stanford University, Stanford, CA 94305, USA

[6]Precourt Institute for Energy, Stanford University, CA 94305, USA

**Abstract**

The design of new applications, especially those based on heterogeneous integration, must rely on detailed knowledge of material properties, such as thermal conductivity (TC). To this end, multiple methods have been developed to study TC as a function of vibrational frequency. Here, we compare three spectral TC methods based on velocity decomposition in homogenous molecular dynamics simulations: Green-Kubo modal analysis (GKMA), the spectral heat current (SHC) method, and a method we propose called homogeneous nonequilibrium modal analysis (HNEMA). First, we derive a convenient per-atom virial expression for systems described by general many-body potentials, enabling compact representations of the heat current, each velocity decomposition method, and other related quantities. Next, we evaluate each method by calculating the spectral TC for carbon nanotubes, graphene, and silicon. We show that each method qualitatively agrees except at optical phonon frequencies, where a combination of mismatched eigenvectors and a large density of states produces artificial TC peaks for modal analysis methods. Our calculations also show that the HNEMA and SHC methods converge much faster than the GKMA method, with the SHC method being the most computationally efficient. Finally, we demonstrate that our single-GPU modal analysis implementation in GPUMD (Graphics Processing Units Molecular Dynamics) is over 1000 times faster than the existing LAMMPS (Large-scale Atomic/Molecular Massively Parallel Simulator) implementation on one CPU.

*Contact: epop@stanford.edu

## I. INTRODUCTION

Advances in the semiconductor industry increasingly require solutions based on new materials as opposed to traditional geometric scaling of silicon transistors. Famous examples are the replacement of $SiO_2$ gate dielectrics with higher-permittivity $HfO_2$ [1], or the use of epitaxial $Si_xGe_{1-x}$ or nitride capping layers to induce strain and improve the channel mobility [2]. This trend will only intensify as, for example, next-generation resistive switching memories may use materials not traditionally found in a semiconductor stack [3] and efforts to develop three-dimensional (3D) heterogeneous integrated circuits may include low-dimensional materials such as two-dimensional (2D) semiconductors or carbon nanotubes (CNTs) [4]. While each technology promises electrical performance improvements, a higher density of interfaces, inclusion of more polycrystalline or amorphous materials, and more nanoscale features will create significant thermal challenges, requiring advanced modeling techniques for thermal characterization.

Proper thermal modeling of any system invariably relies on knowing the thermal conductivity (TC) of its materials. For bulk-like, pristine, crystalline materials, the Peierls-Boltzmann transport equation, with force constants derived from quantum-mechanical density functional theory, can accurately calculate TCs [5]. However, in systems with broken-symmetry, due to defects, amorphization, or contact with other materials, the calculation of TC can be difficult or intractable with quantum-based methods [6]. Molecular dynamics (MD) circumvents symmetry problems, enabling TC calculations of amorphous, polycrystalline, and other exotic structures [6-8]. However, MD's additional flexibility comes with a trade-off, as detailed TC information, such as frequency-dependent (i.e. spectral) contributions, can be difficult to extract.

Spectral TC contributions in solids have been obtained with non-equilibrium MD (NEMD) simulations [9-12] but suffer from size and boundary effects inherent to the NEMD method. Equilibrium MD (EMD) simulations offer some advantages, needing only to fit relevant phonon wavelengths in the simulation cell, because periodic boundaries eliminate boundary scattering and preserve mean free paths. EMD can be used to extract both normal mode relaxation times and phonon dispersions. These data can be used with the Peierls-Boltzmann transport equation to calculate a spectral TC by time domain normal mode analysis [13-16]. The spectral energy density method, which implements normal mode analysis in the frequency domain, is also an EMD method used to gain phonon-level insights for TC [17-22].

Another EMD method, based on the Green-Kubo relations [23], uses the heat flux autocorrelation to calculate the total TC of a system. Lv and Henry combined the Green-Kubo method with normal mode analysis to create Green-Kubo modal analysis (GKMA) [6,24], which decomposes atomic velocities into their modal contributions, enabling modal heat flux and spectral TC calculations. The Green-Kubo relations can also be used in conjunction with small driving forces that enable direct and efficient calculations of the TC. This technique was first proposed by Evans *et al*. [25] but was recently generalized for many-body potentials [26] and is known as the homogeneous nonequilibrium MD (HNEMD) method. Here, the term "homogeneous" is used to denote the lack of a temperature gradient, analogous to EMD methods.

In this work, we combine the HNEMD method with normal mode analysis to create a method called homogeneous nonequilibrium modal analysis (HNEMA). The HNEMA method leverages the modal velocity decomposition from the GKMA method and the efficiency of the HNEMD method to calculate spectral TC. We evaluate and compare the HNEMA method with GKMA as well as the spectral heat current (SHC) method, which uses a virial-velocity correlation function to spectrally decompose TC. First, we derive a per-atom virial tensor, which enables compact representations of each method as well as other related quantities. All three approaches are then applied to calculate the spectral TC of CNTs, graphene, and silicon (Si), comparing the accuracy and uncertainty of each. We also show that the performance of our GPU-implemented algorithms are orders of magnitude faster than CPU-based code.

## II. THEORY

### A. Heat Current and Driving Force in Terms of Per-Atom Virial

In this section, we derive an expression for the per-atom virial, which can be used to unify the calculations of the heat current, driving force, and related quantities. The expression also enables compact representations of various methods, including the GKMA, HNEMA, and SHC that we detail below. The following definitions are valid for general, many-body potentials and have been implemented in the GPUMD (Graphic Processing Units Molecular Dynamics) software [9,26-29].

The total virial tensor for periodic systems can be written as a summation over each atomic virial

$$\mathbf{W} = \sum_i \mathbf{W}_i, \tag{1}$$

where $i$ is the atom index. The atomic virial was originally derived to be [29]

$$\mathbf{W}_i^{orig} = -\frac{1}{2} \sum_{j \neq i} \vec{r}_{ij} \otimes \vec{F}_{ij} \tag{2}$$

with

$$\vec{r}_{ij} \equiv \vec{r}_j - \vec{r}_i, \tag{3}$$

being the position difference vector between atom $j$ and $i$. The expression for the force vector is [29]

$$\vec{F}_{ij} = \frac{\partial U_i}{\partial \vec{r}_{ij}} - \frac{\partial U_j}{\partial \vec{r}_{ji}}, \tag{4}$$

with $U_i$ the potential energy of atom $i$. With the appropriate substitutions, we write the total virial tensor as

$$\mathbf{W} = -\frac{1}{2} \sum_i \sum_{j \neq i} \vec{r}_{ij} \otimes \left( \frac{\partial U_i}{\partial \vec{r}_{ij}} - \frac{\partial U_j}{\partial \vec{r}_{ji}} \right). \tag{5}$$

Manipulating the atom indices, we can show that

$$-\frac{1}{2} \sum_i \sum_{j \neq i} \vec{r}_{ij} \otimes \left( \frac{\partial U_i}{\partial \vec{r}_{ij}} \right) = -\frac{1}{2} \sum_j \sum_{i \neq j} \vec{r}_{ji} \otimes \left( \frac{\partial U_j}{\partial \vec{r}_{ji}} \right) = \frac{1}{2} \sum_i \sum_{j \neq i} \vec{r}_{ij} \otimes \left( \frac{\partial U_j}{\partial \vec{r}_{ji}} \right), \tag{6}$$

which can be used to rewrite Eq. (5) as

$$\mathbf{W} = \sum_i \sum_{j \neq i} \vec{r}_{ij} \otimes \frac{\partial U_j}{\partial \vec{r}_{ji}}. \tag{7}$$

The total virial tensor can again be written as a summation of atomic virials, like in Eq. (1), but the atomic virial we propose is now defined as

$$\mathbf{W}_i^{prop} = \sum_{j \neq i} \vec{r}_{ij} \otimes \frac{\partial U_j}{\partial \vec{r}_{ji}}. \tag{8}$$

It is important to note that the atomic virials in Eqs. (2) and (8) are *not equivalent* (i.e. $\mathbf{W}_i^{orig} \neq \mathbf{W}_i^{prop}$); however, both definitions give the same total virial for the system. For simplicity, we refer to $\mathbf{W}_i^{prop}$ as $\mathbf{W}_i$ for the remainder of the paper.

The advantage of the new per-atom virial definition is that the heat current and related quantities can be conveniently calculated based on it. Previously, the potential part of the per-atom heat current, which is the only part relevant for thermal transport in solids, was written in the following form [29]

$$\vec{J}_i^{\text{pot}} = \sum_{j \neq i} \vec{r}_{ij} \left( \frac{\partial U_j}{\partial \vec{r}_{ji}} \cdot \vec{v}_i \right), \tag{9}$$

where $\vec{v}_i$ is the velocity of atom *i*. Expressing the per-atom heat current in terms of the per-atom virial, we now get

$$\vec{J}_i^{\text{pot}} = \mathbf{W}_i \cdot \vec{v}_i, \tag{10}$$

where the per-atom virial tensor cannot be assumed to be symmetric and the full tensor with nine components should be used. In Ref. [29], it was stated that the potential part of the heat current can only be expressed in terms of the per-atom virial for two-body potentials; however, with our updated per-atom virial, the compact heat current expression of Eq. (10) is valid for general many-body potentials. Note that the form of the heat current allows for it to be easily decomposed [9]. For example, the heat current in the *x*-direction can be decomposed into contributions from atoms' *x*-, *y*-, and *z*-direction motion.

The per-atom virial can also be used to rewrite expressions for the HNEMD method [26]. The HNEMD method provides the framework on which the HNEMA and the SHC methods are built upon. The HNEMD method adds an external driving force to each atom to establish a homogeneous, but nonequilibrium heat current. The driving force was first written [26] in the form

$$\vec{F}_i^{\text{ext}} = E_i \vec{F}_e + \sum_{j \neq i} \left( \frac{\partial U_j}{\partial \vec{r}_{ji}} \otimes \vec{r}_{ij} \right) \cdot \vec{F}_e. \tag{11}$$

Using the definition of the per-atom virial tensor in Eq. (8), we can rewrite it as

$$\vec{F}_i^{\text{ext}} = E_i \vec{F}_e + \vec{F}_e \cdot \mathbf{W}_i. \tag{12}$$

Here, $E_i$ is the total energy of atom *i* and $\vec{F}_e$ is the driving force parameter (in units of inverse length), which must be small enough in magnitude to keep the system within the linear response regime. The expression given in Eq. (12) is implemented in GPUMD [28] and used for all HNEMA and SHC calculations.

**B. Modal Analysis (MA)**

The modal analysis (MA) framework is the foundation for both the GKMA and HNEMA methods. It was first proposed in Ref. [6], detailed further in Ref. [24], and will be briefly reviewed here. A solid with *N* atoms vibrating about their equilibrium positions will have 3*N* vibrational modes [30]. With lattice dynamics (LD), these 3*N* vibrational modes (eigenmodes) can be determined and their polarization vectors (eigenvectors) used as a basis on which to project the anharmonic trajectory realized during an MD simulation. LD may treat the entire simulation cell as a unit cell (i.e. $k = 0 = \Gamma$), which will work for all solids, but any valid basis such as those used for phonon dispersions in crystals, can be used. As shown in Ref. [24], the projection of each atom's velocity coordinates onto the normal mode coordinates should be used (as opposed to force or position coordinates). The transformation from atomic velocity $\vec{v}_i(t)$ of atom *i* to the normal mode velocity $\dot{\vec{X}}_n(t)$ of mode *n*, as well as its inverse transformation, is written as [24]:

$$\dot{\vec{X}}_n(t) = \sum_i \sqrt{m_i} \vec{e}_{i,n}^* \cdot \vec{v}_i(t) \tag{13}$$

$$\vec{v}_i(t) = \frac{1}{\sqrt{m_i}} \sum_n \vec{e}_{i,n} \cdot \vec{\dot{X}}_n(t). \tag{14}$$

Here, $m_i$ is the mass of atom $i$ and $\vec{e}_{i,n}$ is the eigenvector for atom $i$ and eigenmode $n$. The set of eigenvectors $\vec{e}_{i,n}$ used to calculate the modal velocities and heat currents in this work are found using harmonic LD implemented in the `phonon` executable within the GPUMD package [28]. Using these expressions, we can rewrite the heat current in terms of the modal velocity

$$\begin{aligned}\vec{J}^{\text{pot}} &= \sum_i \mathbf{W}_i \cdot \left[ \frac{1}{\sqrt{m_i}} \sum_n \vec{e}_{i,n} \cdot \vec{\dot{X}}_n(t) \right] \\ &= \sum_n \left( \sum_i \frac{1}{\sqrt{m_i}} \mathbf{W}_i \cdot \vec{e}_{i,n} \right) \cdot \vec{\dot{X}}_n(t).\end{aligned} \tag{15}$$

This means that the modal heat current can be written as

$$\vec{J}_n^{\text{pot}} = \left( \sum_i \frac{1}{\sqrt{m_i}} \mathbf{W}_i \cdot \vec{e}_{i,n} \right) \cdot \vec{\dot{X}}_n(t). \tag{16}$$

As shown in the next two subsections, this modal heat current expression can be used with both the Green-Kubo and HNEMD methods to calculate the spectral TC.

**i. Green-Kubo Modal Analysis (GKMA)**

The GKMA method, as described in [6,24], is created simply by substituting the modal heat current definition into the Green-Kubo expression for TC. The modal TC is

$$\kappa_n^{\mu\nu}(\tau) = \frac{1}{k_B T^2 V} \int_0^\tau dt' \langle J_n^\mu(t') J^\nu(0) \rangle \tag{17}$$

with $k_B$ as the Boltzmann constant, $V$ as the simulation cell volume, $T$ is the absolute temperature, and $\tau$ is the integration upper limit of the correlation time. The superscripts $\mu$, $\nu$ are placeholders for the principal $x$, $y$, and $z$ axes. The total heat current is $J^\nu(0)$ and the mode-specific heat current is $J_n^\mu(t')$ with a separation of time $t'$ between samples. The $\langle ... \rangle$ brackets indicate averaging over different time origins. Equation (17) will yield the total contribution to TC of a specific normal mode. More detailed information about mode-mode interactions can be calculated through cross-correlations of modal heat currents [6,24] but, for this work, we will focus on each mode's total contribution to TC. The total TC of the system is the sum of individual mode contributions

$$\kappa^{\mu\nu}(t) = \sum_n \kappa_n^{\mu\nu}(t). \tag{18}$$

In practice, we find the sum of the modes from the GKMA method to match the TC of an independent Green-Kubo method calculation in the same simulation, verifying Eq. (18).

**ii. Homogeneous Nonequilibrium Modal Analysis (HNEMA)**

The HNEMA method only requires a straightforward substitution of the modal heat current into the main expression of TC for the HNEMD method [26]. The resulting expression of the modal TC for the HNEMA method is

$$\frac{\left\langle J_n^\mu(t)\right\rangle_{\text{ne}}}{TV} = \sum_\nu \kappa_n^{\mu\nu} F_e^\nu. \tag{19}$$

Here, $\kappa_n^{\mu\nu}$ is the TC tensor for mode $n$, $T$ and $V$ are the system temperature and volume, respectively, $\left\langle J_n^\mu(t)\right\rangle_{\text{ne}}$ is the mode-specific nonequilibrium heat current, and $\vec{F}_e$ is the driving force parameter. The $\mu$ and $\nu$ terms are, again, placeholders for the principal $x$, $y$, and $z$ axes. Similar to the GKMA method, Equation (18) applies to the HNEMA method, and, in practice, we find the sum of modal TC contributions to match the total TC from an independent HNEMD calculation in the same simulation.

**C. Spectral Heat Current (SHC)**

The SHC method spectrally decomposes the TC of a homogeneous nonequilibrium heat current by first calculating the steady-state time correlation function [26]

$$\vec{K}(t) = \sum_i \sum_{j\neq i} \vec{r}_{ij}(0) \left\langle \left( \frac{\partial U_j}{\partial \vec{r}_{ji}}(0) \cdot \frac{\vec{p}_i(t)}{m_i} \right) \right\rangle_{\text{ne}} \tag{20}$$

with $\vec{p}_i = m_i \vec{v}_i$ being the atomic momentum and $m_i$ the mass of atom $i$. Using our expression for the virial tensor, we can rewrite this as

$$\vec{K}(t) = \sum_i \left\langle \mathbf{W}_i(0) \cdot \vec{v}_i(t) \right\rangle, \tag{21}$$

which reduces to the nonequilibrium heat current (Eq. (10)) when $t = 0$. The remaining derivation follows Ref. [26] with

$$\tilde{\vec{K}}(\omega) = \int_{-\infty}^\infty \vec{K}(t) e^{i\omega t} dt \tag{22}$$

and

$$\kappa(\omega) = \frac{2\tilde{\vec{K}}(\omega)}{VTF_e}. \tag{23}$$

Finally, each of the methods' resulting mode- or frequency-dependent TCs can be further decomposed into contributions from each Cartesian direction [9,26]. For two-dimensional (2D) materials this means that the in-plane and out-of-plane contributions can be distinguished, as we will show for our graphene simulations. This technique has already provided additional physical insight for multiple 2D materials [9,31-33].

**III. VALIDATION AND BENCHMARKS**

**A. Molecular Dynamics Simulations**

The GKMA, HNEMA, and SHC methods are all implemented in GPUMD-v2.5.1 [28]. We compare the TC results and performance of each method for three different crystalline systems: quasi-one-dimensional CNT with a (10,10) chirality, 2D graphene, and 3D Si. Periodic boundary conditions are applied in the axial direction ($x$ direction) for the CNT, in the planar directions ($xy$ plane) for graphene, and in all directions for silicon. For graphene and CNT simulations, we used the Tersoff potential parameterized by Lindsay *et al.* [34] and for Si we used the mini-Tersoff potential [35] (implemented in GPUMD). Every simulation uses a time step of 1 fs and a temperature of 300 K. All results from each method, unless otherwise specified, are aggregated from $N_s = 200$ independent simulations, with a production time of $t_s = 20$ ns, for a total production simulation time of $N_s t_s = 4000$ ns.

The MA methods require a set of eigenvectors to describe vibrations in each material. These eigenvectors must be calculated for the exact structure to be simulated to ensure accurate modal velocity and subsequent spectral TC calculations. To do this, we first relax each of the CNT, graphene, and Si structures at 300 K and zero pressure for 5 ns. From the last 1 ns, we average the cell dimensions to determine the final cell size. Using that final cell size, we minimize the energy of the structure and calculate the eigenvectors and vibrational (phonon) frequencies with harmonic LD using the `phonon` executable within the GPUMD package [28]. The MA methods use the same set of eigenvectors for each material. A histogram showing the count of LD-calculated vibrational frequencies for each material is shown in Fig. 1. We choose to bin the spectral TC contributions based on the vibrational frequency binning used in those histograms. The carbon nanotube and graphene use a bin size of 1 THz for a total of 52 bins, and silicon uses a bin size of 0.25 THz for a total of 63 bins. As a sanity check, we confirmed that the histograms match separate vibrational density of states (VDOS) computations, which are calculated using the velocity autocorrelation function [36], for the same systems.

For all materials, the simulation protocol for the GKMA method is as follows: First, each system is equilibrated in the constant atom number, volume, and temperature ensemble (*NVT*) using the Nosé-Hoover chain thermostat [23] for 100 ps at 300 K. The thermostat is removed, and the system is run in the constant atom number, volume, and energy ensemble (*NVE*) for 100 ps. With the GKMA calculations enabled, the system runs in the *NVE* ensemble for 20 ns. The output data is the modal heat current, which can then be used with Eq. (17) to calculate the modal TC.

Both the HNEMA and SHC methods, which use the nonequilibrium heat current from the HNEMD method, follow the same simulation protocol. First, each system is run for 100 ps in the *NVT* ensemble to equilibrate the temperature to 300 K. Next, either the HNEMA or SHC methods are enabled, both of which apply a small driving force to each atom, and the systems run in the *NVT* ensemble for 20 ns. We note that the HNEMA and SHC methods can be run simultaneously, but we ran them independently for performance measurements. The HNEMA method directly outputs the modal TC [see Eq. (19)] and the SHC method outputs a spectral thermal conductivity [see Eq. (23)].

**i. Results for CNT**

For CNT simulations, we consider a nanotube with $N$ = 8000 atoms, of ~50 nm length and (10,10) chirality. We sample the heat current every 8 fs (124 THz) to avoid aliasing during the GKMA and HNEMA runs, although we find that a sampling rate of 10 fs (100 THz) yields the same results. We use a sampling rate of 2 fs for the SHC method with a maximum correlation time of 500 fs. The per-mode TC for the GKMA method was extracted by averaging over the $\tau$ = [5 ns, 6 ns] range. Finally, our tests show the CNT system is in the linear response regime when the driving force parameter $\left|\vec{F}_\mathrm{e}\right|$ = 0.02 μm$^{-1}$, and we used this value for both the HNEMA and SHC methods.

The results for the CNT spectral TC calculations can be seen in Fig. 2. The total TC (sum or integration over all modes/frequencies) at 300 K is 2091 ± 167 Wm$^{-1}$K$^{-1}$ for the GKMA method, 2132 ± 21 Wm$^{-1}$K$^{-1}$ for the HNEMA method, and 2103 ± 20 Wm$^{-1}$K$^{-1}$ for the SHC method. The spectral TCs of the methods qualitatively agree except for a notable peak at ~49 THz in the MA methods' results. This origin of this peak will be discussed later. There is also qualitative agreement with previous CNT spectral thermal conductivity calculations from Sääskilahti *et al*. [10] for frequencies ≥ 12 THz. They used an NEMD-based method, which requires very large simulation cells for fully diffusive thermal transport. The periodic boundary conditions used for the GKMA, HNEMA, and SHC methods allow for long wavelength phonons (< 12 THz) to scatter appropriately at reasonable cell sizes. We expect Sääskilahti *et al*.'s [10] distributions to match ours as their CNT approaches length $L$ = ∞.

**ii. Results for Graphene**

For graphene simulations, we use a sheet comprised of $N$ = 9072 atoms, of 3.35 Å thickness and ~15×15 nm$^2$ area. We use the same heat current sampling rate as the CNT, 8 fs (124 THz), for the GKMA and

8HNEMA runs. We also use a sampling rate of 2 fs for the SHC method with a maximum correlation time of 500 fs. The per-mode TC for the GKMA method was extracted by averaging over the $\tau = $ [3 ns, 4 ns] range. Since GKMA is an equilibrium method and the TC of graphene is isotropic [37], we can average the $x$- and $y$-direction TC to utilize all data for the GKMA calculations; i.e. $\kappa_{\text{total}} = (\kappa_x + \kappa_y)/2$, resulting in 400 averaged correlations for 200 simulations. The HNEMA and SHC methods rely on a directional nonequilibrium heat current ($x$-direction here) and cannot be averaged as such. We determine the appropriate driving force parameter for graphene to be $|\vec{F}_e| = 0.03$ μm$^{-1}$.

The results for the graphene spectral TC calculations are shown in Fig. 3. Note that the graphene TC has been decomposed into its out-of-plane (flexural) phonon and in-plane phonon modes [26] in addition to the spectral decomposition. The total TC at 300 K is $3073 \pm 129$ Wm$^{-1}$K$^{-1}$ for GKMA, $2928 \pm 17$ Wm$^{-1}$K$^{-1}$ for HNEMA, and $2951 \pm 16$ Wm$^{-1}$K$^{-1}$ for SHC. These results are consistent with previous EMD simulations, which used the same potential and calculated the thermal conductivity to be $2900 \pm 100$ Wm$^{-1}$K$^{-1}$ [9]. Again, all methods qualitatively agree except for the peak at ~49 THz present for the MA subplots. Note that the error bars in Fig. 3(a) for the GKMA method are much larger than the other methods, despite its additional $x$- and $y$-direction averaging.

### iii. Results for Silicon

We use a block of natural Si made of $N = 8000$ atoms, with ~5.4×5.4×5.4 nm$^3$ volume. The Si masses are chosen in accordance with the natural proportions of each isotope (i.e. 92.2 % $^{28}$Si, 4.7 % $^{29}$Si, 3.1 % $^{30}$Si). We sample the heat current every 10 fs (100 THz) for the MA runs, which is much higher than the Nyquist rate of ~31 THz. Like the other materials, we use a sampling rate of 2 fs for the SHC method with a maximum correlation time of 500 fs. The per-mode TC for GKMA was extracted by averaging over the $\tau = $ [2.5 ns, 3 ns] range. While the thermal conductivity of bulk Si is not isotropic, with values between directions varying by 15%-30% [16], we still choose to average the GKMA results over the $x$-, $y$-, and $z$-directions to good effect here; i.e. $\kappa_{\text{total}} = (\kappa_x + \kappa_y + \kappa_z)/3$, resulting in 600 averaged correlations for 200 simulations. As in graphene, we only use $x$-direction data for the HNEMA and SHC methods, and a driving force parameter of $|\vec{F}_e| = 0.1$ μm$^{-1}$ for those simulations.

The spectral TC of Si for all methods is shown in Fig. 4. The calculated total TCs for GKMA, HNEMA, and SHC methods are $150 \pm 4.9$ Wm$^{-1}$K$^{-1}$, $147.3 \pm 0.9$ Wm$^{-1}$K$^{-1}$, and $148.3 \pm 0.8$ Wm$^{-1}$K$^{-1}$, respectively. These values are in good agreement with each other and with the mini-Tersoff development work [35]. On a per-frequency basis, there are significant differences between the results from MA methods and the SHC method. First, the MA results in Fig. 4(a),(b) show multiple bins below 4 THz with no contribution to TC, whereas the SHC results in Fig. 4(c) show meaningful contributions. The SHC method has contributions in that range because, due to the Fourier transform in Eq. (22), it has access to an infinite number of phonon frequencies and is not limited to those resolved by lattice dynamics [11]. In contrast, Fig. 1(c) shows that there are several frequency bins with no vibrational modes, meaning that the LD-based MA methods cannot have TC contributions in those ranges. However, an infinite-size Si crystal would have vibrational modes at those frequencies (see VDOS curve in Fig. 1(c)), and we expect MA and SHC methods to match better as system size increases. [Interestingly, Fig. 5(b) shows that the cumulative TC curves of the MA and SHC methods still overlap well within that frequency range.] The other significant difference between methods is the presence of large peaks in the spectral TC at ~15 THz for the MA methods. These differences will be discussed in the next section.

### B. Error Analysis

In this section, we discuss the accuracy and precision of each method. For accuracy, we will consider which methods obtain the correct spectral TC distribution, and, for precision, we consider how quickly each method converges to its final answer. On a per-frequency basis, there are distinct differences between the MA and the SHC spectral TC results, especially at high frequencies corresponding to the optical phonon branches. The SHC method has minor contributions from the highest frequency phonons, whereas the MA methods exhibit spectral TC peaks in the same range, a difference most notable for silicon. These differences imply methodological or implementation inaccuracies as each method is tested on the same set of structures under the same conditions. To better understand the differences, we examine HNEMA results under different conditions for silicon. We find the high-frequency TC peaks are not due to temperature, system size effects, modal heat current sample rate, numerical precision, changes in mass (i.e. changes in isotopes), choice of thermostat, or choice of driving force parameters.

A major contributing factor to the high-frequency spectral TC peaks is a mismatch between the LD-calculated eigenvectors and those present during MD simulations. If a mismatch is responsible for changes in spectral TC, it is reasonable to expect those changes to be more evident for silicon based on VDOS alone. In Fig. 1, the optical phonon VDOS peak is more pronounced for silicon, with ~40% of the total vibrational modes within 14–16 THz, than graphene and CNT, which have only ~20% within the 48–52 THz range. This expectation also assumes a mismatch results in a heat current that is projected more evenly among the LD-calculated eigenvectors. We test this assumption by considering an extreme eigenvector mismatch and calculate the HNEMA spectral TC for a diamond structure while using a set of silicon eigenvectors.

First, we calculate the eigenvectors for a 4096 atom, ~4.4×4.4×4.4 nm$^3$ block of isotopically pure silicon using the `phonon` executable within the GPUMD package [28] and the mini-Tersoff potential [35]. These eigenvectors are then used as an input to GPUMD for a set of HNEMA simulations; however, we change the input structure to a 4096 atom, ~2.9×2.9×2.9 nm$^3$, diamond structure. We follow the same HNEMA simulation protocol as above, use a driving force parameter of $|\vec{F}_e| = 0.1$ μm$^{-1}$, use the Tersoff potential [38], and run a total of $N_s = 20$ simulations. Figure 5(a) shows that the resulting distributions of the normalized diamond spectral TC and silicon VDOS are qualitatively similar, confirming the assumption that mismatched eigenvectors result in a heat current that is projected more evenly on LD-calculated eigenvectors. Note that low frequency (i.e. < 3 THz) vibrational modes still contribute more to TC on a per-mode basis as larger components of those Si and diamond eigenvectors are likely similar. Overall, this test provides strong evidence that the high-frequency spectral TC peaks in the MA methods are due to a mismatch between LD-calculated and MD-simulated eigenvectors. Finally, it is important to note the high-frequency spectral TC peaks are of similar magnitude for each material, suggesting that the silicon peaks may only look worse because its total TC is much smaller than that of graphene and CNTs.

Comparing to literature and limiting the discussion to silicon, we find the high frequency spectral TC peak is present for other methods as well, although to a lesser degree. The comparison of the normalized cumulative spectral TC for different methods is shown in Fig. 5(b). For the HNEMA method, we find ~20% of silicon's TC comes from optical modes (> 12 THz) with ~13% between 15–16 THz. Zhou et al. [12] used the time domain direct decomposition method (TDDDM), the time domain normal mode analysis method, and the spectral energy density method, which all resulted in ~10% of the Si thermal conductivity coming from the optical modes. Each of these methods is based on LD calculations, but extra emphasis was placed on quasi-harmonic LD calculations and achieving small temperature fluctuations (i.e. < 0.1 K; smaller than intrinsic fluctuations), which improved their agreement between LD-calculated and MD-simulated eigenvectors [12]. In their following work, they calculated spectral TC using the frequency domain direct decomposition method (FDDDM), which had ~15% of the total TC coming from optical modes [11]. The FDDDM does not use LD and a larger optical mode contribution cannot be explained by an eigenvector mismatch. The stochastically initialized temperature-dependent effective potential (s-TDEP) method, which is based on density functional theory, has also been used to calculate the spectral TC [39]; however, we find that the optical mode contributions better match those from the SHC method, comprising ~5% of the total thermal conductivity. While the s-TDEP calculation does not include all orders of anharmonicity, since we are only considering a temperature of 300 K, higher order phonon interactions are

10not crucial [40]. Finally, we note that contributions to TC from optical modes are expected, but to what extent seems to depend on different factors such as method, system size, etc.

From a precision perspective, knowing the rate of convergence is important when considering different spectral TC methods. Here, we use the standard error $\text{SE} = \sigma/\sqrt{n}$ as a metric for precision, with $\sigma$ and $n$ denoting the standard deviation and number of simulations, respectively. We consider the maximum SE ($\text{SE}_{\text{Max}}$) of all frequency bins (or equivalent frequency ranges for SHC) in a material as we increase the number of simulations. This choice is arbitrary; however, other precision metrics should result in the same conclusions. The $\text{SE}_{\text{Max}}$ results can be seen in Fig. 6. Note that the HNEMA method's $\text{SE}_{\text{Max}}$ is much smaller than that of the GKMA method. This is because there is a decreasing signal-to-noise ratio in the heat current autocorrelation function used for any Green-Kubo based method, making accurate property extractions difficult at long correlation times [26]. This is not the case for the HNEMA method, which has a constant signal-to-noise ratio due to its direct calculation of thermal conductivity. As a result, the $\text{SE}_{\text{Max}}$ for 200 simulations of the GKMA method can be matched by fewer than 20 simulations of the HNEMA method for each of the materials studied here.

The SHC method, which also uses a correlation function, should suffer from similar convergence issues as the GKMA method; however, we find its $\text{SE}_{\text{Max}}$ to be comparable to that of the HNEMA method. This is because the SHC method only needs correlation times long enough to resolve phonon frequencies (i.e. < 1 ps), whereas the GKMA method needs correlation times that can capture phonon relaxations (i.e. up to many nanoseconds for crystalline materials). The GKMA method, with its long correlation times, is subject to more noise and requires significantly more simulations to converge than the HNEMA or SHC methods. For this study of $\text{SE}_{\text{Max}}$, each frequency bin consists of many vibrational modes. If individual vibrational modes need to be considered, the variance will be much larger and may render the GKMA computationally impractical. However, other storage, memory, and computational constraints must be considered. These will be discussed in the next section.

### C. Performance Analysis

**i. Method Comparison**

In practice, users may need to choose a method based on their available computational resources. Here, we consider costs based on the implementation of each method in GPUMD [28]. (A detailed examination of implementation and computational cost for GKMA can be found in Ref. [24].) First, the MA methods require LD calculations which have time and space complexities of $O(N^3)$ and $O(N^2)$, respectively. These calculations run once per structure and, given that many MA simulations are required to converge to the final answer, generally will not dominate the total computation time. The `phonon` executable in the GPUMD package can be used to accelerate these calculations, but memory is currently limited to that available on a single NVIDIA GPU. As the LD-calculated eigenvectors must be loaded to run MA MD simulations, the MA space complexity is also $O(N^2)$. Both the GKMA and HNEMA methods are based on the modal heat current calculation in Eq. (16) and have a time complexity of $O(N^2)$ if calculating all the modes. If needed, both time and space complexity can be reduced to $O(MN)$, where $M$ is a subset of the $N$ total modes.

The GKMA is a postprocessing method, meaning that a large amount of space is required to save the raw data. Careful choices of frequency bins and smaller heat current sampling frequencies can substantially reduce the space requirement. Despite our binning and sampling choices, which reduced storage requirements by over a factor of 1000, the GKMA output files in this work were still ~20 gigabytes (GB) per simulation. With large output files, which totaled ~4 terabytes per material in this work, calculating Eq. (17) can be cumbersome. Because the HNEMA method directly computes (and averages) the modal thermal conductivity, the heat current sampling rate and file writes can be decoupled. For example, our graphene

HNEMA simulations output modal thermal conductivity every $10^4$ time steps resulting in output files 1250 times smaller than GKMA. However, although there are steep storage requirements for the GKMA method, the availability of the modal heat current can be used for mode-mode cross-correlations [6]. These calculations cannot be done with the HNEMA method. Finally, the correlation function in Eq. (17) can be calculated during simulations, which can drastically reduce the GKMA output file size.

The SHC method is not based on LD and does not require an eigenvector input file. Because the maximum correlation time is generally small (i.e. < 1 ps) and the sample period is also small (i.e. 1-10 fs), it is practical to compute the correlation function during simulations. With the correlation calculation included during runtime, the time and space complexity is still linear, but increases to $O(Nn_c)$, where $n_c$ is the number of correlation steps (and is usually constant for a given material). The storage requirements are then reduced to the order of kilobytes. Overall, the SHC method is much faster, less resource intensive, and scales much better than the MA methods. For the simulations presented in this paper, the 20 ns production run was approximately four to five times faster with the SHC method.

## ii. GPUMD and LAMMPS Comparison

While the HNEMA and SHC methods are currently only implemented on GPUMD, a CPU-based GKMA code [24] is available for the LAMMPS package [41-43]. Here, we will compare the results and performance between the GPUMD and LAMMPS implementations of the GKMA method. The heat flux calculations in LAMMPS have been shown to be incorrect [29,44,45], especially for low dimensional materials, so we compare a 3D Si system. The mini-Tersoff potential is implemented in GPUMD only, and we choose to use the original Si Tersoff potential [38] in this section. First, we compare the spectral TC results. For simplicity, we simulate a block of 216 isotopically pure Si atoms, with ~1.6×1.6×1.6 nm$^3$ volume. The simulation protocol is a 100 ps, *NVT* equilibration run at 300 K, followed by another 100 ps *NVE* run, but then followed by an *NVE* production run of 10 ns with GKMA calculations. We use a time step of 1 fs, a bin size of 27 modes-per-bin (as opposed to binning by frequency), and we sample the modal heat current every 10 fs. The eigenvectors are calculated with the GULP package [46], and the modal thermal conductivities were extracted by averaging over the $\tau$ = [0.5 ns, 1 ns] range.

A comparison between the cumulative TC curves calculated by GPUMD and LAMMPS for silicon, averaged over $N_s$ = 50 simulations, is shown in Fig. 7. The curves overlap over the entire frequency range, confirming that both the GPU and CPU implementations are consistent. This also suggests the high frequency peak in spectral TC, at ~15 THz, is not due to an implementation error. As mentioned before, the high frequency peak is likely from a mismatch between the LD-determined vibrational modes and those present at finite temperature. Since the same set of eigenvectors is used in each simulator, any mismatch would present itself identically.

Next, we compare the speed and memory limitations. First, we consider an 8000 atom Si cube. We use the same, original Tersoff potential as above [38], sample the modal heat current every time step, and output three bins (i.e. bin size of 8000). For the GPUMD simulations, we run on two sets of hardware. The first is the GeForce RTX 2080 Ti GPU with 11 GB of on-device memory and the second is the V100 GPU with the SXM2 form factor and 16 GB of on-device memory. The LAMMPS simulations are run on compute nodes from the Sherlock cluster at Stanford University [47]. Each compute node has two Intel E5-2640v4 CPUs with a total of 20 cores, 128 GB of memory, and enhanced data rate (100 Gb/s) InfiniBand interconnects. The LAMMPS package is built with OpenMPI and GCC.

In Fig. 8(a), we see how the CPU-MPI speed scales with number of CPU cores. Extrapolating the single CPU core performance (i.e. ideal scaling), we find that lower limit on the CPU cores needed to match the GPUMD performance is 1802 and 2663 cores for the 2080 Ti and V100 GPUs, respectively. The CPU-MPI code does not scale perfectly and, at 1600 CPU cores, the LAMMPS code is still approximately four times slower than GPUMD on the 2080 Ti. Figure 8(b) shows how the speed of each simulator changes with number of atoms. Note that the current GPUMD implementation can only run on a single GPU and,

12since the MA methods have $O(N^2)$ space complexity, large system simulations (> ~25,000 atoms) are only possible on LAMMPS at this time. Additionally, further optimizations can be made for the LAMMPS code, which may make it a few times faster. In the end, if a user's GPU has enough memory, it will be significantly faster to run the GPUMD implementation.

## IV. CONCLUSIONS

In this work, we have compared three homogenous velocity decomposition methods used for spectral thermal conductivity (TC) calculations: GKMA, SHC, and HNEMA, the latter of which we have proposed here. We derived a per-atom virial which enabled compact representations and unified expressions for heat current, driving force, and each of the spectral TC methods. We evaluated each by simulating CNT, graphene, and Si systems and found that modal analysis (MA) results qualitatively agree with SHC, except at optical phonon frequencies. Further investigation showed that LD eigenvector mismatches, due to finite temperature MD simulations, combined with high densities of optical modes, resulted in artificial contributions to spectral TC. These effects were small for graphene and CNTs, but significant for silicon. The spectral TC simulations also showed the HNEMA and SHC to be approximately one to two orders of magnitude more efficient than the GKMA method with respect to total simulation time. From a performance perspective, both MA methods' computation and memory requirements scale quadratically with system size compared to linear scaling for the SHC method. As a result, the SHC method can be significantly cheaper to use. Finally, we compared our GPU-implemented MA code with available LAMMPS CPU code and demonstrated, first, that each produced the same results and, second, that a single GPU running GPUMD was over 1000 times faster than one CPU using the current LAMMPS implementation.

**Acknowledgments**

Some of the computing for this project was performed on the Sherlock cluster at Stanford University. We would like to thank Stanford University and the Stanford Research Computing Center (SRCC) for providing computational resources and support that contributed to these results. This work was also partially supported by ASCENT, one of the six centers in JUMP, a Semiconductor Research Corporation (SRC) program sponsored by DARPA. A.J.G. also acknowledges support from the Achievement Rewards for College Scientists (ARCS) Northern California Chapter. T.A-N. has been supported in part by the Academy of Finland through its QTF Center of Excellence (project no. 312298). Z.F. has been supported by the National Natural Science Foundation of China (NSFC) (No. 11974059).

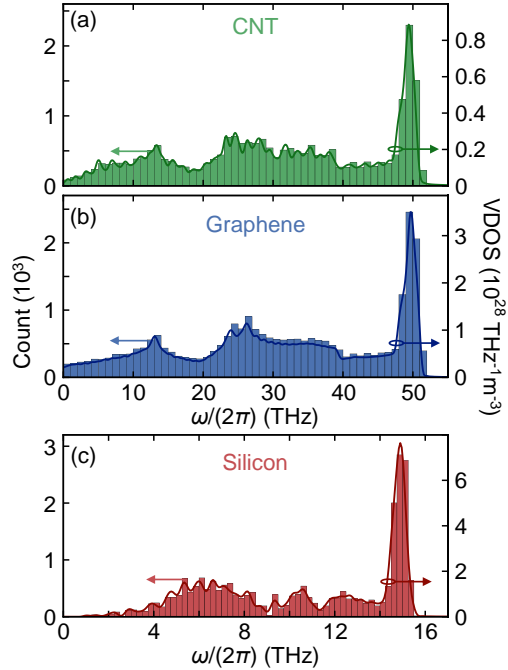

**FIG. 1**: Histogram of the count of vibrational frequencies (left *y*-axis, bars) from LD calculations and the vibrational density of states (VDOS; right *y*-axis, lines) for our (a) CNT, (b) 2D graphene, and (c) 3D natural silicon systems. Note the horizontal frequency axis is the same for CNT and graphene, but different for silicon.

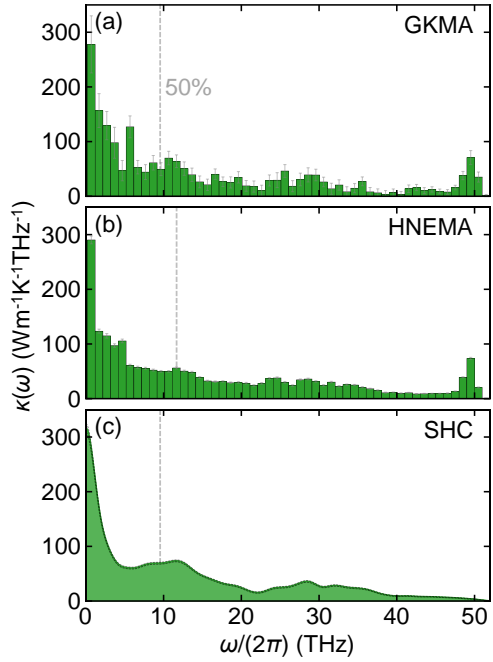

**FIG. 2**: Spectral decomposition of thermal conductivity for a ~50 nm long, (10,10) CNT using (a) the GKMA method, (b) the HNEMA method, and (c) the SHC method. The light gray error bars in (a),(b), and the range between dashed lines in (c) show the standard error. The gray vertical dashed lines denote the frequency below which 50% of the total TC is contributed.

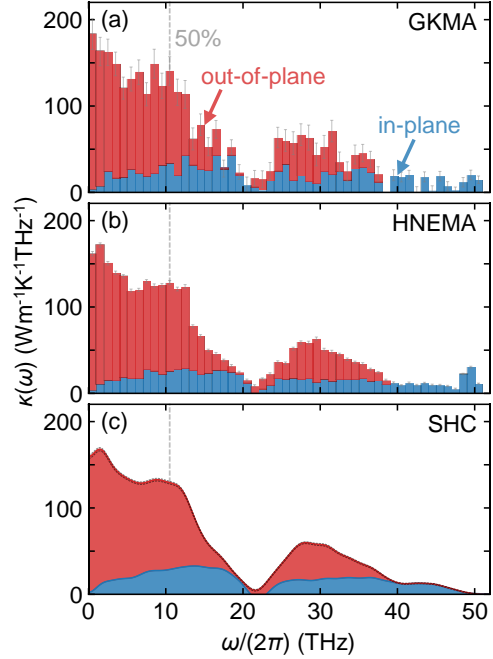

**FIG. 3**: Spectral decomposition of thermal conductivity for a ~15×15 nm² single-layer graphene sheet using (a) the GKMA method, (b) the HNEMA method, and (c) the SHC method. The light gray error bars in (a), (b), and the range between dashed lines in (c) show the standard error. The red regions denote the contributions to thermal conductivity from the out-of-plane (flexural) phonon modes and the blue regions denote those from in-plane phonons. Each subplot shows a stacked-style bar or line plot. The gray vertical dashed lines denote the frequency below which 50% of the total TC is contributed.

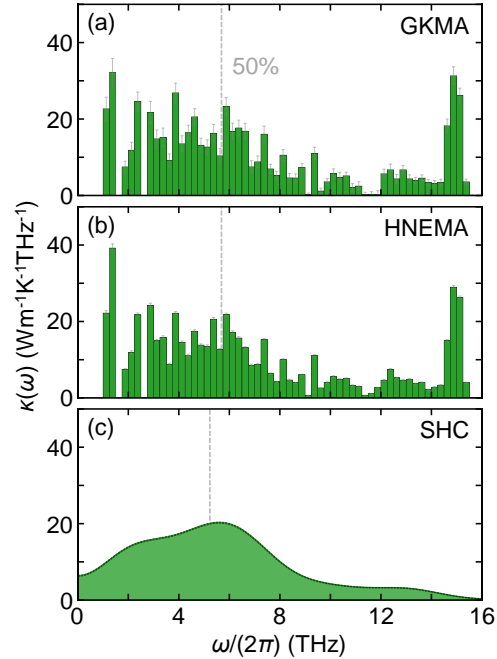

**FIG. 4**: Spectral decomposition of thermal conductivity for a ~5.4×5.4×5.4 nm³ cube of natural silicon using the (a) GKMA, (b) HNEMA, and (c) SHC method. Light gray error bars in (a), (b), and the range between dashed lines in (c) show the standard error. The gray vertical dashed lines denote the frequency below which 50% of the total TC is contributed.

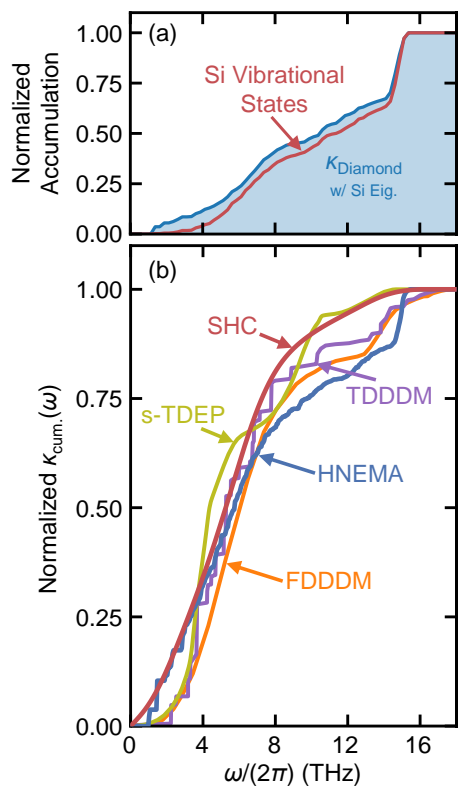

**FIG. 5**: (a) Normalized cumulative curves for the vibrational modes of a 4096-atom block of natural Si as well as the HNEMA spectral TC of a diamond structure using those Si modes. (b) Comparison of methods for the cumulative thermal conductivity of Si with respect to frequency. The TDDDM [12] and FDDDM [11] methods are based on NEMD simulations, the HNEMA and SHC methods are based on HNEMD simulations, and the s-TDEP [39] method is based on anharmonic phonon density functional theory calculations. The GKMA method effectively overlaps with the HNEMA curve, so it is not shown here. The spectral energy density and time domain normal mode analysis methods outputs from Zhou *et al.* [12] overlap the TDDDM curve and are not shown here as well.

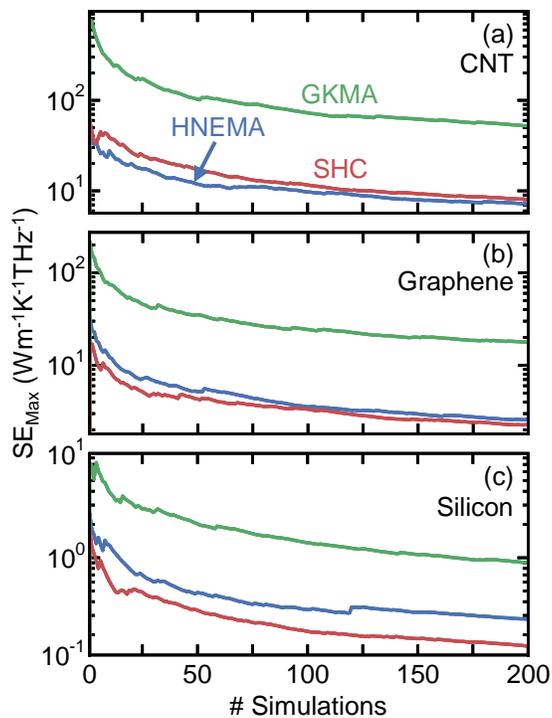

**FIG. 6**: The maximum standard error (SE$_{Max}$) in the spectral thermal conductivity calculation of our (a) CNT, (b) graphene, (c) natural silicon as a function of the number of simulations. The SE$_{Max}$ is calculated for a 1 THz bin or range for the CNT and graphene, and a 0.25 THz bin or range for Si. The green lines indicate SE$_{Max}$ for the GKMA method, blue for HNEMA, and red for SHC.

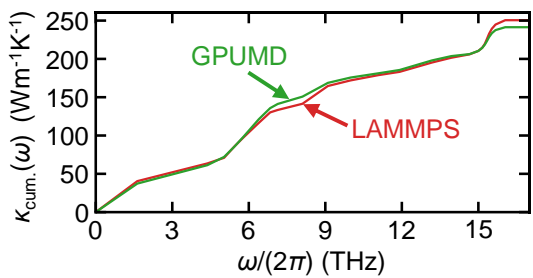

**FIG. 7**: Cumulative thermal conductivity of isotopically pure silicon vs. frequency for the GKMA method run in GPUMD (green) and LAMMPS (red).

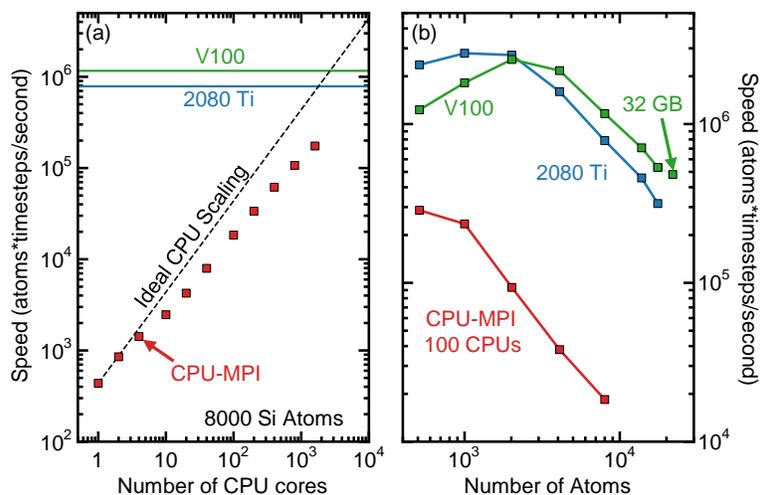

**FIG 8**: (a) The speed of the GPUMD and LAMMPS simulators for GKMA simulations of an 8000 atom isotopically pure Si system. GPUMD simulations are run with 1 CPU on the V100 (green line) and 2080 Ti (blue line) GPUs. Red squares denote LAMMPS simulation performance. The dashed black line is a linear scaling of the speed of a 1 CPU LAMMPS run. (b) GPUMD and LAMMPS speeds with number of atoms at fixed resources. For the largest structure of 21,952 Si atoms, the 32 GB variant of the V100 was used.